\documentclass{IEEEtran}  

\usepackage{import}
\usepackage[utf8]{inputenc}
\usepackage[footnotes,definitionLists,hashEnumerators,smartEllipses,hybrid]{markdown}
\usepackage[capitalise,noabbrev,nameinlink]{cleveref}
\usepackage{graphicx}
\usepackage{adjustbox}
\usepackage{epstopdf}
\usepackage[labelformat=simple]{subcaption}

\usepackage{siunitx}
\usepackage{tabularx}
\usepackage{booktabs}
\usepackage[para,online,flushleft]{threeparttable}
\usepackage{colortbl}

\usepackage{fp}
\usepackage{tikz}
\usetikzlibrary{external}
\usepackage{tikzscale}
\usepackage{pgfplots}

\usepgfplotslibrary{groupplots}
\pgfplotsset{width=10cm,compat=1.16}
\usepackage{svg}
\usepackage{todonotes}
\usepackage{xfrac}
\usepackage{comment}
\usepackage[numbers]{natbib}

\definecolor{goodgreen}{HTML}{c5eecc}
\definecolor{goodred}{HTML}{ffc7cc}
\definecolor{bblack}{HTML}{E2DBCC}

\newcommand{\MinNumber}{-1}%
\newcommand{\MaxNumber}{1}%

\newcommand{\ApplyGradient}[1]{%
  \FPeval{\Percent}{100.0*(#1-\MinNumber)/(\MaxNumber-\MinNumber)}%
  \pgfmathsetmacro{\PercentColor}{\Percent}%
  \xdef\PercentColor{\PercentColor}%
  \cellcolor{goodgreen!\PercentColor!goodred}{#1}%
}
\newcommand*{\SetGradientLimits}[2]{%
    \renewcommand*{\MinNumber}{#1}%
    \renewcommand*{\MaxNumber}{#2}%
}
\newcolumntype{H}{>{\collectcell\ApplyGradient}X<{\endcollectcell}}

\title{\huge
Computational Analysis of Subscapularis Tears and \\Pectoralis Major Transfers on Muscular Activity}

\author{Fabien P\'{e}an, Philippe Favre, Orcun Goksel
\thanks{F.~P\'{e}an and O.~Goksel are with the Computer-assisted Applications in Medicine (CAiM) group, ETH Zurich, Switzerland.}
\thanks{P.~Favre is with Zimmer Biomet, Winterthur, Switzerland.
}}

\begin{document}
\maketitle

\begin{abstract}
Muscle transfers are commonly performed to restore joint function after muscle tears. 
However, there are not many biomechanical studies of muscle transfers, with the available ones often limited to passive movements in anatomical planes.
Using data from three activities of daily living (ADL) and an available computational musculoskeletal model of the shoulder, we analyse the impact of a subscapularis tear and its treatment by a pectoralis major (PMA) transfer of the clavicular, sternal, or both segments.
The shoulder model is validated against experimental data: the model kinematics with motion capture, muscle activity with EMG measurements, and model joint reaction force with in-vivo data from an instrumented prosthesis.
Our results indicate that subscapularis tear requires a compensatory activation of the supraspinatus and is accompanied by a reduced co-contraction of the infraspinatus, both of which can be partially recovered after PMA transfer.
Furthermore, although the PMA acts asynchronously to the subscapularis before the transfer, its patterns of activation change significantly after the transfer.
Overall, this study demonstrates that muscle transfers have a significant impact on the neuromuscular system of the joint during active motion, beyond mere kinematics. Capability of a transferred muscle segment to activate similarly to intact subscapularis is found to be dependent on a considered motion. Differences in the activation patterns between intact subscapularis and segments of PMA may explain the difficulty of some patients in adapting their psycho-motor patterns during the rehabilitation period. Thereby, rehabilitation programs could benefit from targeted training on specific motion patterns and biofeedback programs.
Finally, the condition of the anterior deltoid should be considered to avoid adding limitations to the joint function before a transfer of the clavicular part. \emph{Keywords} --- Musculoskeletal simulation, Shoulder, Upper extremity, Muscle transfer, Activities of daily living, Muscle activation, Muscle recruitment, Rehabilitation
\end{abstract}

\section{Introduction}
A torn rotator cuff muscle may severely impair the mobility of the shoulder, resulting in  restricted range-of-motion, pain, difficulties to perform common daily life activities and reduced quality of life~\cite{Gumina2017clinical, Minagawa2013}.
Irreparable tears can be treated with a muscle transfer, where the insertion site of a functioning muscle is detached and transferred onto the insertion area of the torn muscle.
The aim of such procedure is to restore the function of the torn muscle, while maintaining the original function carried out by the transferred muscle~\cite{Elhassan2010,Axe2016,Clark2018}. 

The subscapularis muscle is the largest muscle of the rotator cuff, also the only one with a relatively anterior position~\cite{Keating1993}. 
Subscapularis tears have been increasingly reported, owing to imaging improvements and better awareness~\cite{Gerber1991,Gerber1996,Lee2018}, with muscle transfer being a common treatment to restore the subscapularis function.
Since the initial study by Wirth et Rockwood~\cite{Wirth1997}, the latissimus dorsi, teres major~\cite{Elhassan2014}, and pectoralis minor \cite{Wirth1997,Paladini2013} have all been considered for muscle transfer for subscapularis tears. The most common choice, however, is the pectoralis major (PMA), with the options being a transfer of the entire muscle~\cite{Jost2003}, the sternal part only~\cite{Elhassan2008,Valenti2011,Valenti2015}, or the clavicular part only\cite{Resch2000,Gavriilidis2010,Valenti2011,Valenti2015}.
Although PMA transfer has been shown to be less favorable in patients suffering from massive rotator cuff tears~\cite{Jost2003,Nelson2014,Shin2016}, positive outcomes of PMA transfer  include reduced pain~\cite{Resch2000,Gavriilidis2010,Moroder2017,Shin2016}, improved activities of daily living (ADL)~\cite{Jost2003,Gavriilidis2010}, higher Constant score~\cite{Resch2000,Jost2003,Elhassan2008,Shin2016,Ernstbrunner2019}, and a better joint stability~\cite{Wirth1997,Resch2000}.
The choice of which muscle to transfer is typically based on anatomical considerations, with the literature consisting mainly of retrospective clinical studies assessing the efficacy of different transfers. 
Ex-vivo biomechanical experiments have been performed to quantify the changes in moment arms~\cite{Favre2008a,Ackland2008,Hik2019}. However, such studies focus on the function of only the transferred muscle, and neglect the complex active muscle interplay of the entire shoulder girdle.

Numerical models can simulate muscle activity during motion. Effects of various clinical procedures can then be tested in a reproducible fashion on the same  anatomy, e.g.\ allowing for a direct comparison of different muscle transfers.
Previous biomechanical studies of muscle transfers have mainly focused on the transfer of the latissimus dorsi or the teres major for supero-posterior rotator cuff tears~\cite{Magermans2004biomechanical,Magermans2004effectiveness,Favre2008a}, while PMA transfer studies were limited to in-vitro kinematics or in-silico static analysis of simple motions~\cite{Konrad2007,Jastifer2012}.
In this work, the functional outcome of PMA transfer for isolated tears of the subscapularis is evaluated during activities of daily living, using a multi-body, muscle-driven, finite element model of the shoulder~\cite{Pean2020surface}.
Tear of the subscapularis was simulated and the transfer of different segments (clavicular, sternal, or both parts) of the PMA was investigated.
Simulated muscle activation of the rotator cuff, deltoid anterior, and pectoralis major parts were assessed for three ADL (eating, washing, and combing).

\section{Methods}
\subsection{Musculoskeletal Model of the Shoulder}
The biomechanical model used herein was described in detail in previous works~\cite{Pean2020surface,Pean2020rsa}. 
Some of the main features are summarized herein, for sake of completeness.
Our multi-body model involves four rigid structures (thorax, clavicle, scapula, and humerus) and 19 muscles segments representing 13 muscles listed in \cref{tbl:muscle_abbr}.
\begin{table}
\centering
\caption{List of muscle segments in the model with their abbreviations.}
\label{tbl:muscle_abbr}
\begin{adjustbox}{max width=\linewidth}
\begin{tabular}{ll}
\toprule
\textbf{Name}               & \textbf{Abbr.}                        \\ 
\midrule
Deltoid Anterior            & DAN                                   \\ 
Deltoid Middle              & DMI                                   \\ 
Deltoid Posterior           & DPO                                   \\ 
Infraspinatus               & ISP                                   \\ 
Latissimus Dorsi            & LD                                    \\ 
Levator Scapulae            & LS                                    \\ 
Pectoralis Major Abdominal  & PMAA                                  \\ 
Pectoralis Major Clavicular & PMAC                                  \\ 
Pectoralis Major Sternal    & PMAS                                  \\ 
Pectoralis Minor            & PMI                                   \\ 
Rhomboid                    & RM                                    \\ 
Serratus Anterior           & SAN                                   \\ 
Subscapularis               & SSC                                   \\ 
Supraspinatus               & SSP                                   \\ 
Teres Major                 & TMA                                   \\ 
Teres Minor                 & TMI                                   \\ 
Trapezius Inferior          & TRI                                   \\ 
Trapezius Middle            & TRM                                   \\ 
Trapezius Superior          & TRS                                   \\ 
\bottomrule
\end{tabular}
\end{adjustbox}
\end{table}%
The muscle segments were modelled as membrane elements, automatically meshed with B-spline surfaces as in~\cite{Pean2020surface}.
Using contact constraints, these membrane muscle segments wrap on the bones.
Muscle material is modeled as a combination of an embedding matrix using a co-rotated Hooke’s law and a fiber component with passive and active parts modelled according to Blemker et al.~\cite{Blemker2005a}. 
Detailed shoulder model settings and constraints are presented in~\cite{Pean2020surface}, and the specific simulation parameters used herein are given in~\cite{Pean2020rsa}.
Contraction of the fibers in each muscle part is controlled by a muscle activation parameter, normalized between 0 and~1.
For our functional simulation, the differences in position and velocity between body markers from motion capture and the corresponding landmarks on the bone model are minimized during the motion.

\subsection{Subject-specific Modelling}
Eating, washing, and combing motions from a publicly available shoulder movement dataset~\cite{Bolsterlee2014b} were simulated. 
This dataset contains motion capture of several upper limb landmarks: 4 
on the torso; 4 
on the scapula; 3 
on the clavicle; and 2 
on the humerus (with exact locations defined in~\cite{Wu2005}). These landmarks were also annotated on our computational model.
All motions for all 9 marker locations were transformed to a fixed thorax coordinate frame at all time instances by computing a rigid transformation of the thorax frame to its initial position at $t=0$.
We follow the procedure described in \cite{Pean2020rsa} to register our model to the subject's markers and anthropomorphic measurements.

\subsection{Electromyography (EMG)}
We used the raw EMG measurements from the above-mentioned dataset~\cite{Bolsterlee2014b}, which includes the following muscle parts: Deltoid (anterior, medial, posterior), Infraspinatus, Latissimus Dorsi, Pectoralis Major (clavicular, sternal), Trapezius (inferior, transverse, superior). 
Following~\cite{Staudenmann2007}, the raw EMG data was first filtered by a high-pass 1st order Butterworth filter at 250\,Hz, then rectified, and filtered by a low-pass 1st order Butterworth filter at 2\,Hz. 
The similarity between the model activation prediction and the processed EMG was evaluated using Spearman's $\rho$, yielding a correlation coefficient $x_\text{corr}$ that describes the monotonic relationship between the variables while obviating any parametric model assumption between them.

\subsection{Comparison to in-vivo Joint Reaction Forces}
Simulated joint reaction force (JRF) magnitudes were compared to in-vivo measurements obtained from instrumented prostheses~\cite{Bergmann2007,Westerhoff2009a}. 
Comparison was performed for the combing motion, the only common activity between both datasets.
In our results, four measurements (herein called S1, S2, S3, and S4) are included from the public Orthoload database\footnote{S1, S2, S3, and S4, respectively, correspond to the prosthesis readings \mbox{\emph{s1r\_180805\_1\_36}}, \mbox{\emph{s2r\_150506\_1\_88}}, \mbox{\emph{s3l\_110906\_1\_46}}, and \mbox{\emph{s4r\_130407\_1\_15}} in the Orthoload database: \url{https://orthoload.com/}}.

\subsection{Simulation of Pectoralis Major Transfer}
\begin{figure*}
    \centering
    \begin{subfigure}{0.25\linewidth}
        \includegraphics[width=\linewidth,trim=150 200 100 0,clip]{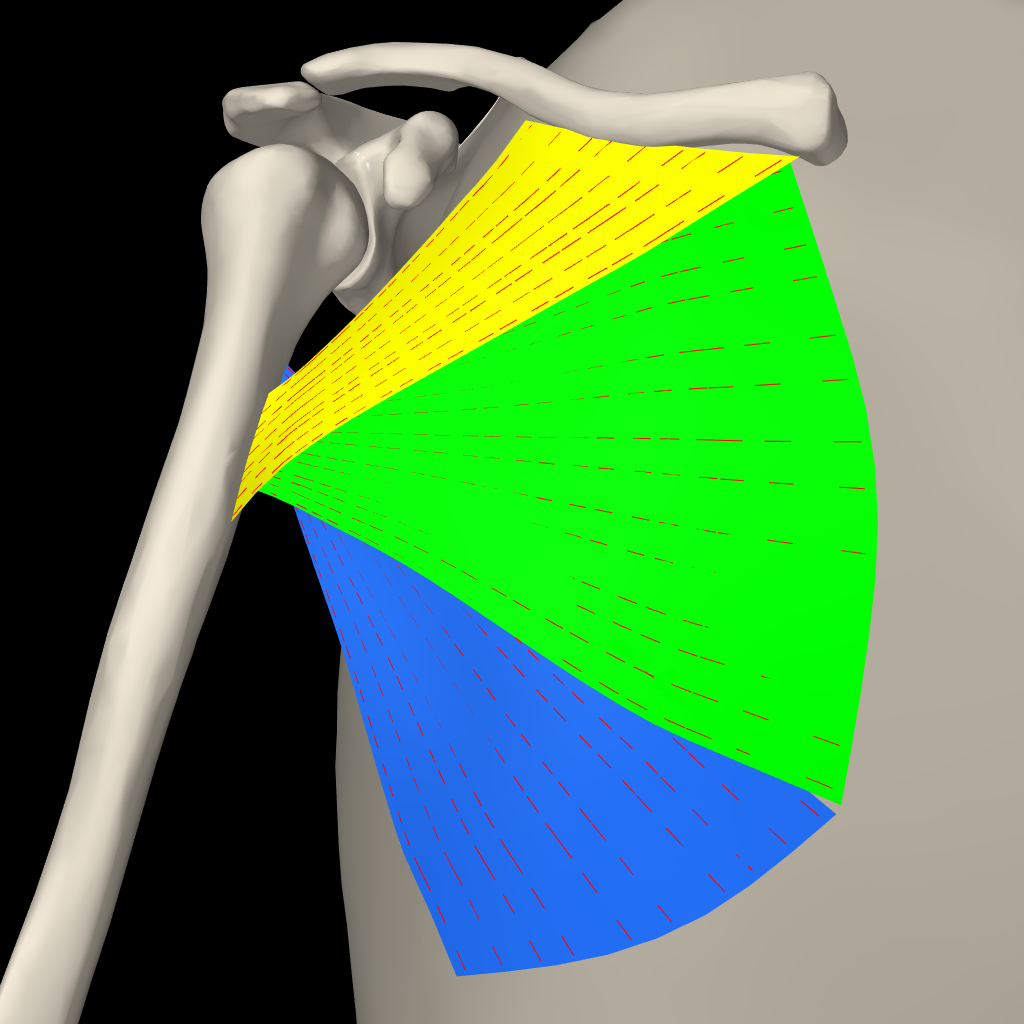}%
        \subcaption{}
        \label{fig:transfer_intact}
    \end{subfigure}%
    \begin{subfigure}{0.25\linewidth}
        \includegraphics[width=\linewidth,trim=150 200 100 0,clip]{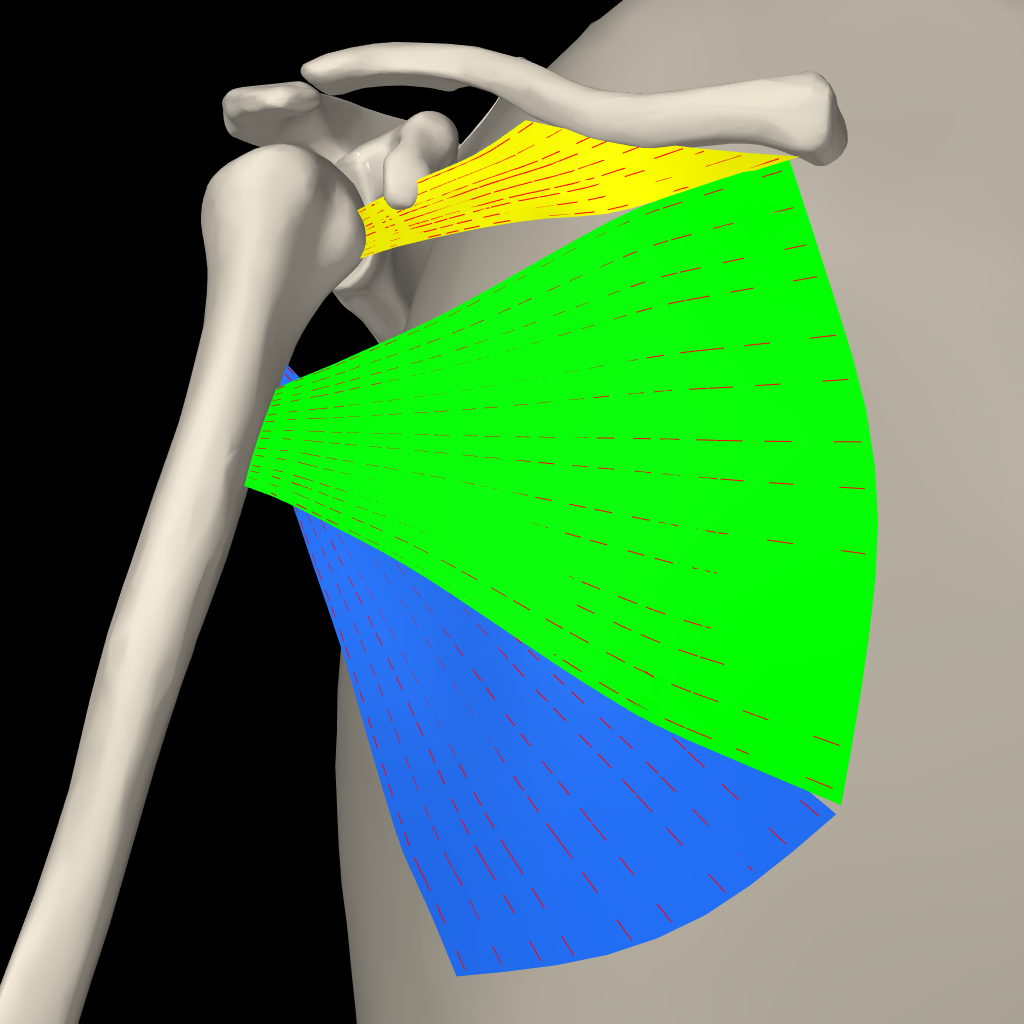}%
        \subcaption{}
        \label{fig:transfer_pmac}
    \end{subfigure}%
    \begin{subfigure}{0.25\linewidth}
        \includegraphics[width=\linewidth,trim=150 200 100 0,clip]{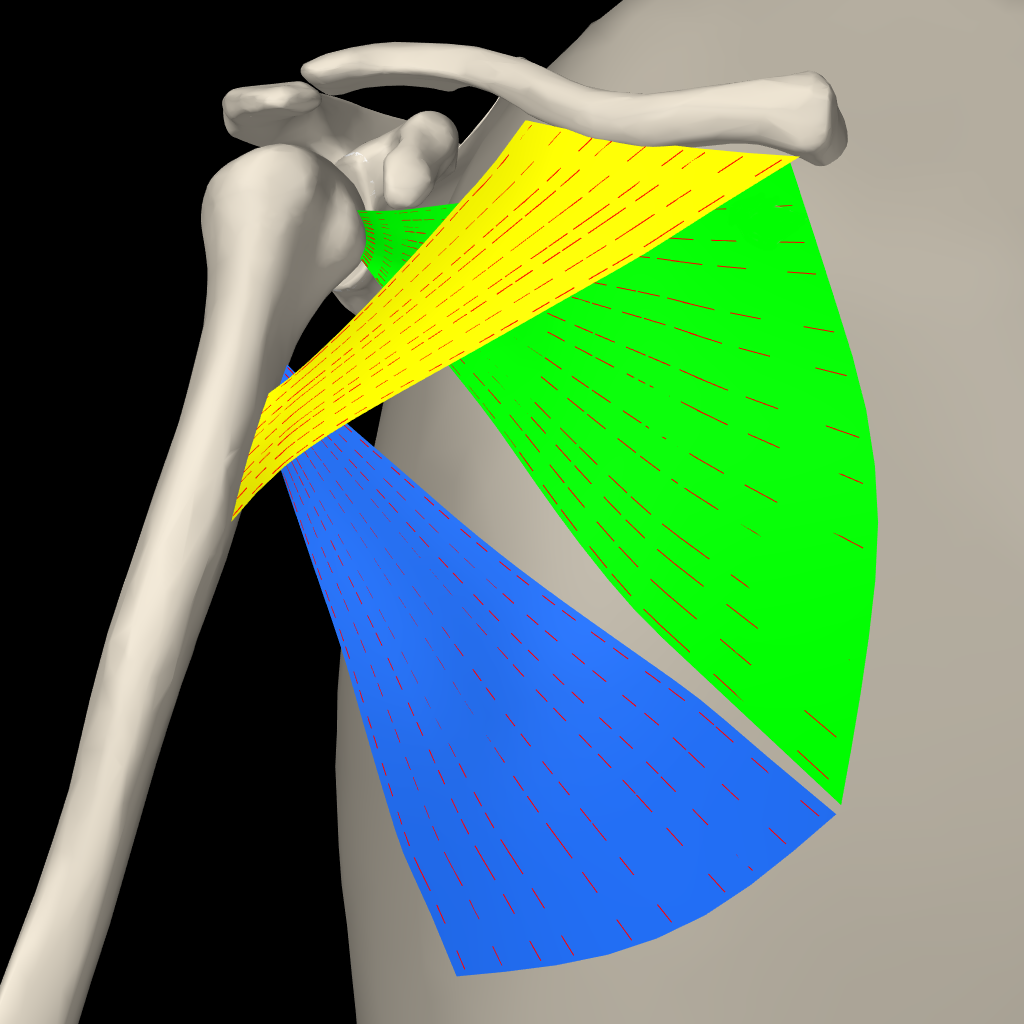}%
        \subcaption{}
        \label{fig:transfer_pmas}
    \end{subfigure}%
    \begin{subfigure}{0.25\linewidth}
        \includegraphics[width=\linewidth,trim=150 200 100 0,clip]{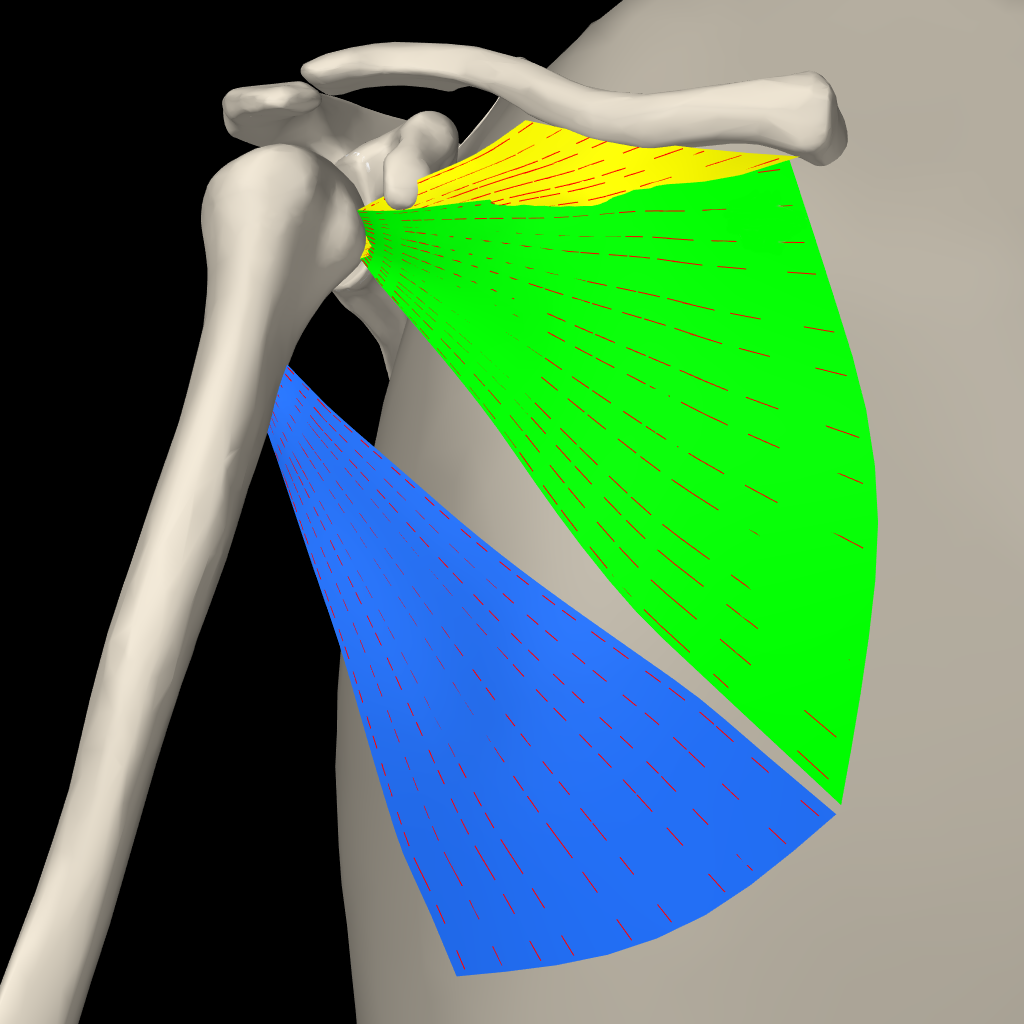}%
        \subcaption{}
        \label{fig:transfer_pmasc}
    \end{subfigure}%
    \caption{Intact PMA~\subref{fig:transfer_intact} simulated transfer of \subref{fig:transfer_pmac} clavicular part, \subref{fig:transfer_pmas} sternal part, and \subref{fig:transfer_pmasc} sternoclavicular part. Pectoralis major clavicular, sternal, abdominal parts are shown in yellow, green, blue, respectively. }
    \label{fig:transfer_visualization}
\end{figure*}

PMA is commonly divided into three parts, the clavicular part (PMAC), the sternal part (PMAS), and the abdominal part, which were represented in our model by three independent muscle segments that can be activated separately. 
In line with previous work~\cite{Shin2016}, transfer of PMAC and PMAS were studied. These were computationally performed by moving their respective insertion sites to the SSC insertion site, see \cref{fig:transfer_visualization}.
\emph{Intact} computational scenario with physiologically functioning SSC was taken as the baseline.
An isolated, irreparable tear of the subscapularis was then simulated by removing the muscle from the simulation, therefore removing both its passive and active force-generating capacity; referred to as case  ``$-$SSC'' with minus representing removal.
Transfers of the clavicular part (+PMAC), sternal part (+PMAS), or both (+PMASC) were then performed on the model.
For all conditions, the following three ADL were simulated: eating, washing axilla, and combing.

\subsection{Analysis of the Tear and the Muscle Transfers}
Muscle transfer procedures aim not only to replace the function of the torn muscle by the transferred muscle, but also to keep the remaining muscles to function as normal as possible, e.g.\ preventing them from  overloading to avoid further tears.  
Specifically, the loss of the physiological contribution of the transferred muscle should not lead to secondary functional impairments.
Accordingly, first we analyzed the impact of the tear and the following transfers on neighbouring muscles (SSP, ISP, and DAN), by comparing their activation levels to the intact case.
Co-contraction of ISP with SSC is evaluated for timesteps where SSC activation is above an arbitrary threshold of 1\,\%  in the intact case. Similarly, changes in the activation signals of PMAC and PMAS were compared with their intact state.

In order to replace the active function of SSC, the transferred muscle should be activated when SSC was itself active in the intact case~\cite{Omid2013,Shin2016}. 
This is evaluated as the percentage of time when PMAC or PMAS is active at the same time as SSC was active in the intact case. In this analysis, SSC, PMAC and PMAS were considered active when activation signals were above an arbitrary threshold of 1\,\%.
In addition, spearman correlation coefficients were computed to evaluate how closely the transferred PMAC or PMAS act in-phase with the intact SSC. 
Spearman correlation coefficients were also computed to assess how in-phase the PMA activation is after its transfer, compared to its intact condition.
Finally, stability of the glenohumeral joint is reported for all cases as the deviation of the JRF from a pure compressive direction. 

\section{Results}
\subsection{Model Validation}
\begin{figure}[t]
    \centering
    \def\svgwidth{\linewidth}
    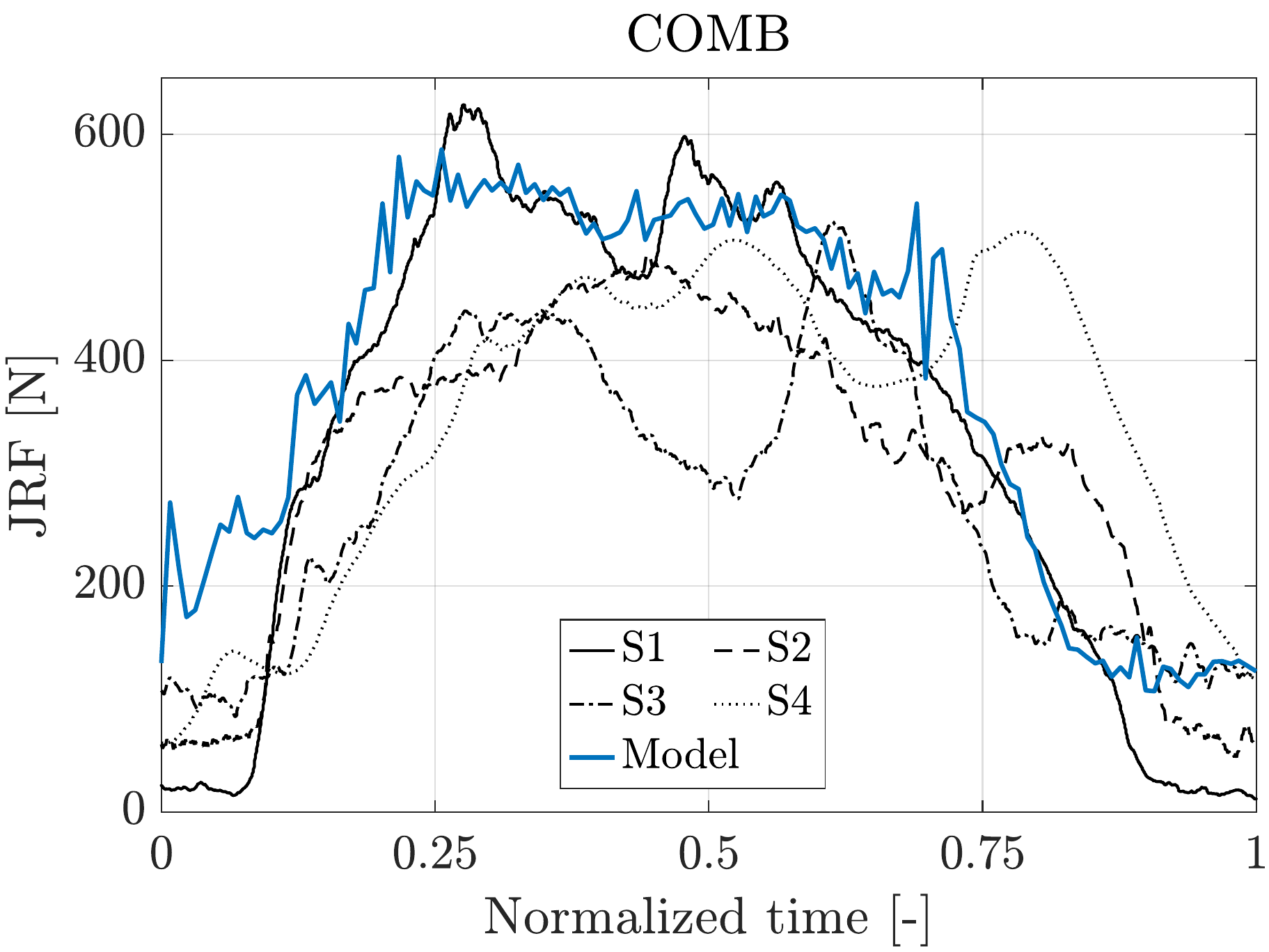
    \caption{Joint Reaction Force magnitude for the combing motion. In-vivo measurements are shown in black and current model results in blue. All curves are time-normalized to the range [0,1].}
    \label{fig:jrf_comb_comparison}
\end{figure}

The three ADL could be successfully simulated by the intact model, with an average distance error  for the humerus landmarks (EM, EL) of 1.2\,cm for eating, 0.7\,cm for washing, and 3.4\,cm for combing (\cref{tab:tracking_quality_intact}.
Landmark positions with a simulated SSC tear lie on average within 1.2\,cm of the intact positions, and improves to below 0.9\,cm after different transfers (see \cref{tab:tracking_quality_condition}).
\begin{table}[t]
\centering
\caption{Distance, mean$\pm$SD~(max), between landmarks located at the humeral tip, EM and EL, and their respective target positions from motion capture data.}
\label{tab:tracking_quality_intact}
\begin{adjustbox}{max width=\linewidth}
\begin{tabular}{lccc}
\toprule
\emph{\small [cm]} & EAT                & WASH               & COMB               \\
\midrule
EM           & $1.2\pm\!0.2\ (1.8)$ & $0.7\pm\!0.2\ (1.4)$ & $3.4\pm\!1.5\ (5.1)$ \\
EL           & $1.1\pm\!0.3\ (1.8)$ & $0.7\pm\!0.3\ (1.7)$ & $1.9\pm\!0.4\ (3.0)$ \\
\bottomrule
\end{tabular}
\end{adjustbox}
\end{table}%
\begin{table}[t]
\centering
\caption{Distance, mean$\pm$SD~(max), between the simulated intact motion and the simulated clinical conditions of the humerus landmarks EM and EL. }
\label{tab:tracking_quality_condition}
\begin{adjustbox}{max width=\linewidth}
\begin{tabular}{lccc}
\toprule
\emph{\small [cm]} & EAT & WASH & COMB \\ \midrule
EM & & & \\
\quad$-$SSC   & $0.6\pm\!0.3\ (1.4)$        & $0.2\pm\!0.1\ (0.5)$         & $1.2\pm\!0.7\ (2.0)$         \\ 
\quad+PMAC  & $0.5\pm\!0.3\ (1.4)$        & $0.2\pm\!0.1\ (0.9)$         & $0.7\pm\!0.4\ (1.6)$         \\ 
\quad+PMAS  & $0.4\pm\!0.3\ (1.1)$        & $0.2\pm\!0.1\ (0.6)$         & $0.9\pm\!0.4\ (1.4)$         \\
\quad+PMASC & $0.4\pm\!0.2\ (1.0)$        & $0.3\pm\!0.2\ (1.1)$         & $0.8\pm\!0.3\ (1.5)$         \\
EL & & & \\
\quad$-$SSC   & $0.6\pm\!0.4\ (1.6)$        & $0.1\pm\!0.1\ (0.3)$         & $0.9\pm\!0.6\ (1.9)$         \\
\quad+PMAC  & $0.5\pm\!0.4\ (1.7)$        & $0.2\pm\!0.1\ (1.0)$         & $0.9\pm\!0.5\ (1.8)$         \\
\quad+PMAS  & $0.4\pm\!0.2\ (1.1)$        & $0.2\pm\!0.1\ (0.6)$         & $0.8\pm\!0.4\ (1.5)$         \\
\quad+PMASC & $0.4\pm\!0.2\ (1.2)$        & $0.3\pm\!0.2\ (1.1)$         & $0.9\pm\!0.3\ (1.7)$         \\
\bottomrule
\end{tabular}
\end{adjustbox}
\end{table}
Simulated activations for the intact case positively correlate ($x_\text{corr}>0.5$) with EMG for at least one motion for all reported parts except PMAS~(\cref{tab:emg_corr}).
\begin{table}[t]
\centering
\SetGradientLimits{-1}{1}
\aboverulesep=0pt
\belowrulesep=0pt
\caption{For the intact scenario, Spearman correlation coefficients $x_\text{corr}$ between simulated activations and the processed EMG signals from the reference dataset. Cells are color coded from -1 in red to 1 in green. Muscle abbreviations are listed in Table \ref{tbl:muscle_abbr}.}
\label{tab:emg_corr}
\begin{tabular}{lrrrrrrr}
\toprule
 & DAN  & DMI  & DPO  & PMAC & PMAS & ISP  \\
\midrule
EAT  &  \ApplyGradient{0.47} & \ApplyGradient{0.41}  & \ApplyGradient{-0.06}   & \ApplyGradient{0.64} & \ApplyGradient{-0.03}     & \ApplyGradient{0.81}\\
WASH & \ApplyGradient{0.78} & \ApplyGradient{-0.30} & \ApplyGradient{0.23} & \ApplyGradient{0.83} & \ApplyGradient{0.25}  & \ApplyGradient{0.36}\\
COMB & \ApplyGradient{0.83} & \ApplyGradient{0.57}  & \ApplyGradient{0.59} & \ApplyGradient{0.26} & \ApplyGradient{-0.12} & \ApplyGradient{0.86}\\
\bottomrule
\end{tabular}
\end{table}
The simulated JRF shows trends and magnitudes similar to the in-vivo data (\cref{fig:jrf_comb_comparison}).

\subsection{Muscle Activity}
\begin{figure}[t]
    \centering
    \def\svgwidth{\linewidth}
    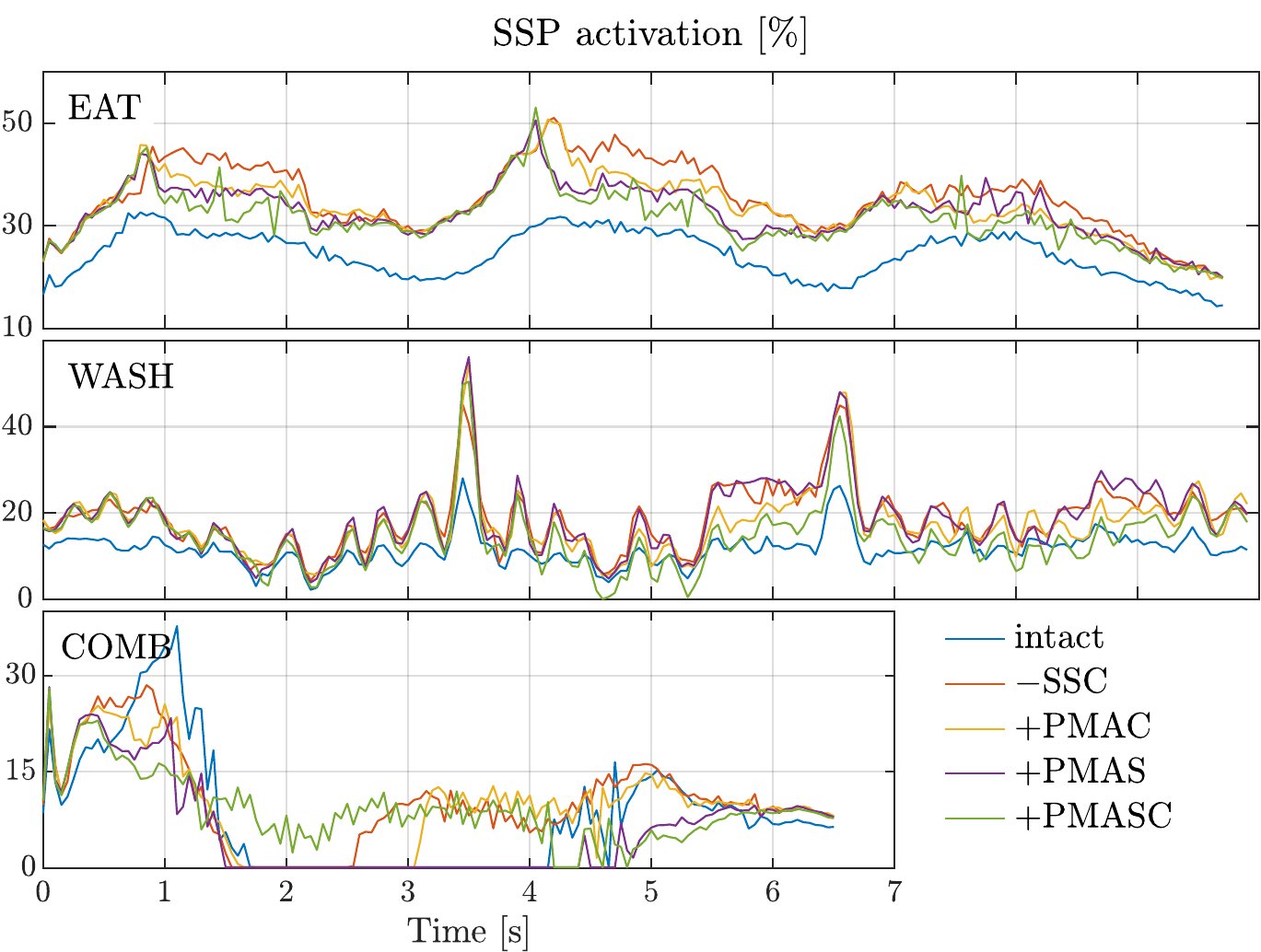
    \caption{Computed activation of the SSP for each simulated ADL for all simulated cases.}
    \label{fig:act_ssp}
    \centering
    \def\svgwidth{\linewidth}
    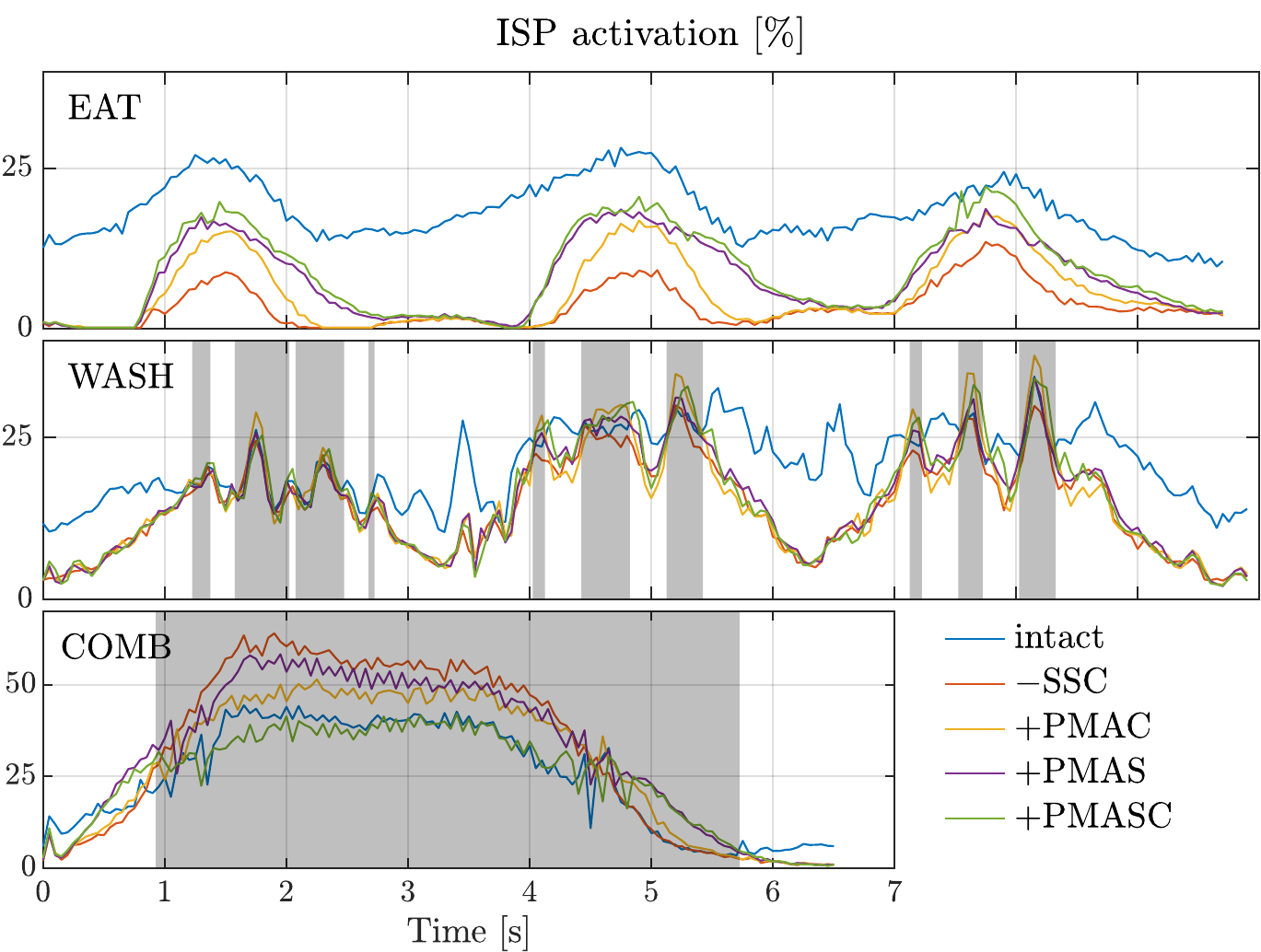
    \caption{Computed activation of the ISP for each simulated ADL for all cases. Shaded areas correspond to time windows where activation of the SSC in the intact case is below a threshold of 1\%, indicating the absence of co-contraction of SSC and ISP.}
    \label{fig:act_isp}
\end{figure}

Activation of the SSP with the torn SSC was consistently higher than in the intact case, with a mean  difference of 11.2, 7.2, and 2.3\,\%, respectively, for the eat, wash, and comb motions. 
Muscle transfers generally reduced the activation of the SSP compared to the torn SSC case (\cref{tab:act_ssp}) 
but not to the level of the intact baseline, as clearly visible for the eating motion (\cref{fig:act_ssp}).  
\begin{table}[t]
\centering
\caption{Mean difference of SSP activation after transfers with respect to its activation in the torn SSC case, i.e.\ $\sum_i^N\frac{1}{N}(a_\text{ssp}^\text{case}(t_i)-a_\text{ssp}^\text{-ssc}(t_i))$, shown in the format mean$\pm$SD.}
\label{tab:act_ssp}
\begin{tabular}{lccc}
\toprule
Case & EAT & WASH & COMB \\ \midrule
+PMAC  & -1.8$\pm$2.4 & -1.2$\pm$3.3 & -0.8$\pm$3.3 \\
+PMAS  & -2.9$\pm$3.2 & \phantom{-}0.0$\pm$2.3  & -4.5$\pm$4.6 \\
+PMASC & -4.1$\pm$4.1 & -4.1$\pm$3.8 & -1.6$\pm$6.0 \\ \bottomrule
\end{tabular}
\end{table}%
\begin{table}[t]
\centering
\caption{Mean difference of ISP activation after transfers with respect to its activation in the torn SSC case, i.e.\ $\sum_i^N\frac{1}{N}(a_\text{isp}^\text{case}(t_i)-a_\text{isp}^\text{-ssc}(t_i))$, shown in the format mean$\pm$SD.}
\label{tab:act_isp}
\begin{tabular}{lccc}
\toprule
Case & EAT & WASH & COMB \\ \midrule
+PMAC  & 2.2$\pm$2.5 & -0.1$\pm$1.7 &  0.7$\pm$1.0 \\
+PMAS  & 4.4$\pm$3.8 &  \phantom{-}0.7$\pm$1.3  & 3.1$\pm$4.3 \\
+PMASC & 5.3$\pm$4.3 & \phantom{-}0.7$\pm$1.9 &   2.7$\pm$3.5 \\ \bottomrule
\end{tabular}
\end{table}

Activation of the ISP in the torn SSC case was lower than in the intact case (\cref{fig:act_isp}).
The mean difference of activation during co-contraction periods are: -14.9, -8.1, and -4.2\,\%, respectively, for the eating, washing, and combing motion.
Muscle transfers generally improved the co-contraction of the ISP compared to the torn SSC case (\cref{tab:act_isp}) 
but not to the level of the intact baseline, as clearly visible for the eating motion (\cref{fig:act_isp}).  

\begin{figure}[t]
    \centering
    \def\svgwidth{\linewidth}
    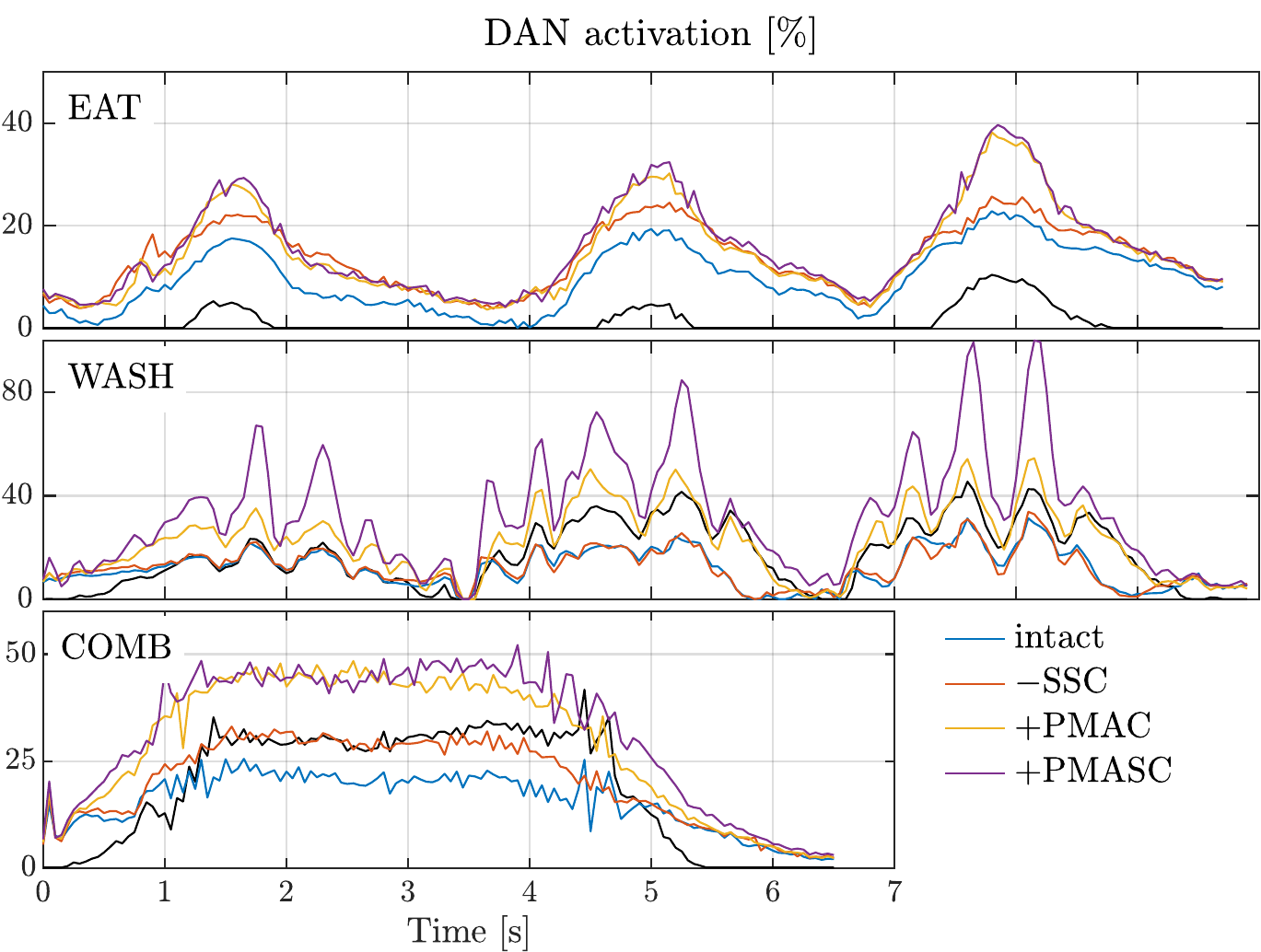
    \caption{Computed activation of the DAN for each simulated ADL for the intact, torn SSC, and transfers involving PMAC cases. Activation of the intact PMAC is shown for comparison (black solid line).}
    \label{fig:act_dan_comparison}
\end{figure}

The activation of the anterior part of the deltoid increased after PMAC transfer (\cref{fig:act_dan_comparison}).
The transfer of the clavicular part resulted in an average increase in DAN activation of 5, 10.3, and 13.5\,\%, while the transfer along with the sternal part increased on average by 6, 20.6, and 16.2\,\% for the eat, wash, and comb motion, respectively.

\begin{figure}[t]
    \centering
    \def\svgwidth{\linewidth}
    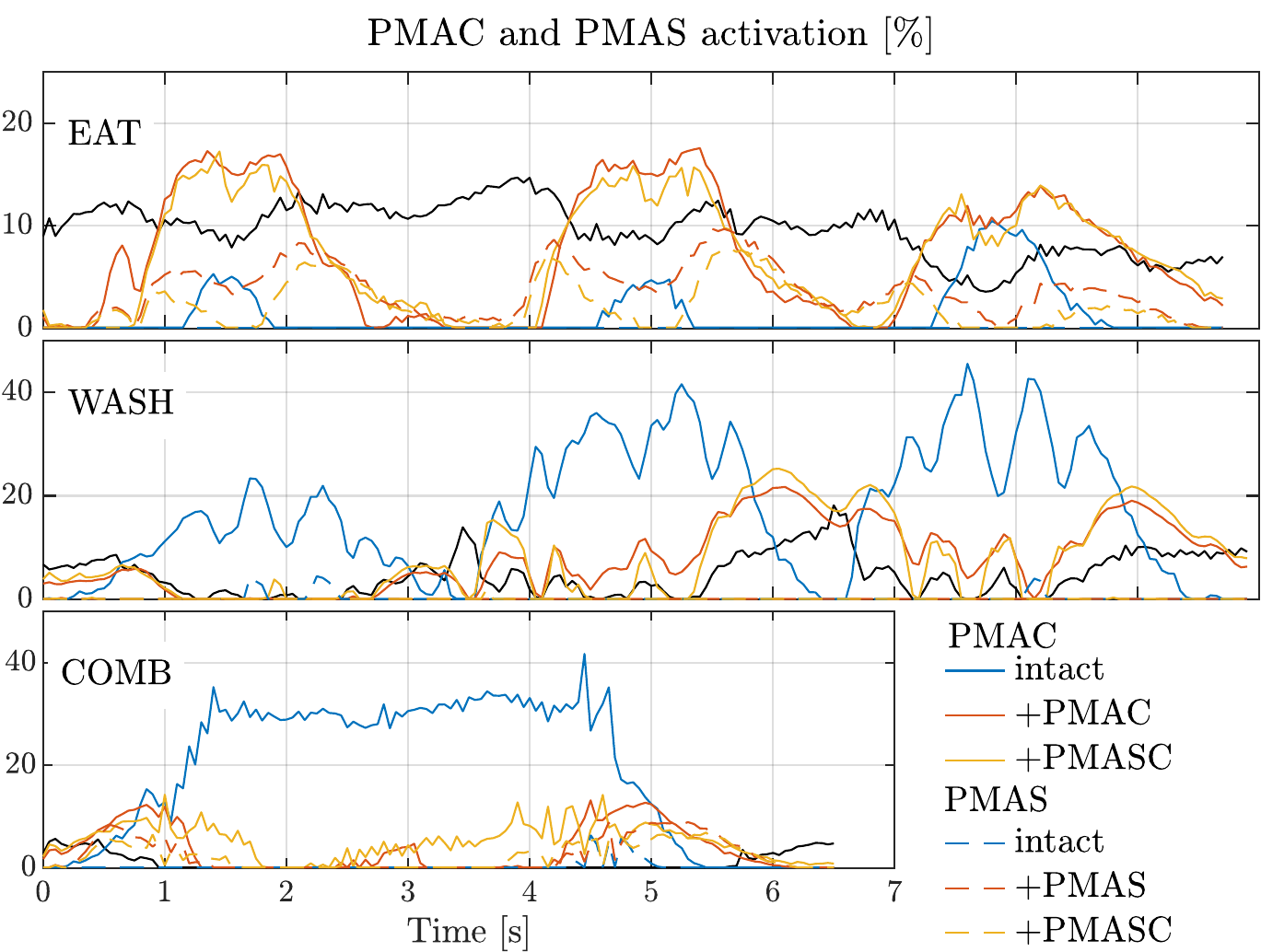
    \caption{Computed activation of the PMAC (solid lines) and PMAS (dashed lines) for each simulated ADL for the intact and torn SSC cases, and transfer cases involving the respective muscle segments transfer. Activation of the intact SSC is shown for comparison (black solid line).}
    \label{fig:act_pma_comparison}
\end{figure}

After their respective transfer, PMAC and PMAS activation coincided more with intact SSC activation for the eating and combing motions (\cref{tab:cocontraction_ssc_pma}). For the washing motion, PMAC was already highly active when the intact SSC was active, and it improved slightly after transfer. However, the opposite was seen with PMAS. 
The activation of PMAC and PMAS in the intact case are negatively correlated with the intact SSC activation for all motions (\cref{fig:act_pma_comparison}). Although no clear trend was observed for the eating and combing motions~($x_\text{corr}\le-0.19$), during washing motion PMAC pattern after transfer correlates positively with the intact SSC~($x_\text{corr}\ge0.64$), see \cref{tab:act_pattern_pma_ssc}.
PMAS, instead, positively correlates after its transfer with the intact SSC for the eating motion~($x_\text{corr}\ge0.21$).
\begin{table}[t]
\caption{Percentage of the time during which the PMAC and PMAS are active ($\ge1\%$) at the same time as SSC was in the intact case. Cells are color coded from 0 in red to 100 in green.}
\label{tab:cocontraction_ssc_pma}
\aboverulesep=0pt
\belowrulesep=0pt
\SetGradientLimits{0}{100}
\centering
\begin{adjustbox}{max width=\linewidth}
\begin{tabular}{lccccccccc}
\toprule
     & \multicolumn{3}{c}{PMAC}               & \phantom{a} & \multicolumn{3}{c}{PMAS}               \\ \cmidrule{2-4} \cmidrule{6-8}
     & intact & +PMAC  & +PMASC & & intact & +PMAS & +PMASC \\ \midrule
EAT  & \ApplyGradient{28}    & \ApplyGradient{77}    & \ApplyGradient{81}   & & \ApplyGradient{00}    &  \ApplyGradient{85}  & \ApplyGradient{62}   \\
WASH & \ApplyGradient{81}    & \ApplyGradient{90}    & \ApplyGradient{93}   & & \ApplyGradient{00}    &  \ApplyGradient{01}  & \ApplyGradient{00}   \\
COMB & \ApplyGradient{40}    & \ApplyGradient{66}    & \ApplyGradient{83}   & & \ApplyGradient{00}    &  \ApplyGradient{54}  & \ApplyGradient{57}   \\
\bottomrule
\end{tabular}
\end{adjustbox}
\end{table}%
\begin{table}[t]
\centering
\aboverulesep=0pt
\belowrulesep=0pt
\SetGradientLimits{-1}{1}
\centering
\caption{Spearman correlation coefficients between the activation signal of SSC in the intact case and the activation signals of PMAC and PMAS in intact case 
and after both transfer. Cells are color coded from -1 in red to 1 in green. Empty cells indicate no correlation ($\rho$=NaN), as PMAS shows no activation in the intact case.}
\label{tab:act_pattern_pma_ssc}
\begin{adjustbox}{max width=\linewidth}
\begin{tabular}{lrrrcrrr}
\toprule
\multicolumn{1}{l}{} & \multicolumn{3}{c}{PMAC} & \phantom{a}& \multicolumn{3}{c}{PMAS}\\ \cmidrule{2-4} \cmidrule{6-8}
\multicolumn{1}{l}{} & intact & +PMAC & +PMASC  &  & intact & +PMAS & +PMASC \\ \midrule
EAT  & \ApplyGradient{-0.55} & \ApplyGradient{-0.38} & \ApplyGradient{-0.41} &  & \cellcolor{bblack}{ }  & \ApplyGradient{0.21}  & \ApplyGradient{0.40}  \\
WASH & \ApplyGradient{-0.64} & \ApplyGradient{0.64}  & \ApplyGradient{0.79}  &  & \ApplyGradient{-0.39}  & \ApplyGradient{0.07} & \ApplyGradient{0.09} \\
COMB & \ApplyGradient{-0.70} & \ApplyGradient{-0.19} & \ApplyGradient{0.11}  &  & \ApplyGradient{-0.20} & \ApplyGradient{0.14}  & \ApplyGradient{0.01} \\
\bottomrule   
\end{tabular}
\end{adjustbox}
\end{table}

With a torn SSC, PMAC and PMAS both activate similarly to their respective intact condition~($x_\text{corr}\ge0.67$), see \cref{tab:act_pattern_pma_pma}.
However, PMAC and PMAS activation patterns mostly changed significantly after their transfer, compared to their intact activations. 
\begin{table}[t]
\centering
\aboverulesep=0pt
\belowrulesep=0pt
\SetGradientLimits{-1}{1}
\centering
\caption{Spearman correlation coefficients between activation signals in the intact case and with torn SSC and after both transfer. Cells are color coded from -1 in red to 1 in green. Empty cells indicate no correlation ($\rho$=NaN), as PMAS shows no activation in the intact case.
}
\label{tab:act_pattern_pma_pma}
\begin{adjustbox}{max width=\linewidth}
\begin{tabular}{lrrrcrrr}
\toprule
     & \multicolumn{3}{c}{PMAC} & \phantom{a} & \multicolumn{3}{c}{PMAS} \\  \cline{2-4} \cmidrule{6-8}
     & $-$SSC  & +PMAC  & +PMASC  &  & $-$SSC  & +PMAS  & +PMASC  \\ \midrule
EAT  & \ApplyGradient{0.96}  & \ApplyGradient{0.62}   & \ApplyGradient{0.63}    &  & \cellcolor{bblack}{ }   & \cellcolor{bblack}{ }    & \cellcolor{bblack}{ }     \\
WASH & \ApplyGradient{0.99}  & \ApplyGradient{0.06}  & \ApplyGradient{-0.27}   &  & \ApplyGradient{0.78}  & \ApplyGradient{-0.04}  & \ApplyGradient{-0.03}   \\
COMB & \ApplyGradient{0.92}  & \ApplyGradient{0.19}   & \ApplyGradient{-0.34}   &  & \ApplyGradient{0.67}  & \ApplyGradient{0.40}   & \ApplyGradient{0.39}    \\
\bottomrule
\end{tabular}
\end{adjustbox}
\end{table}

On average, joint stability was worse when the SSC was torn compared to the intact condition. After PMA transfers, stability was generally slightly improved compared to the torn SSC case (\cref{tab:stability}), but not to the level of the intact SSC condition.
\begin{table}[]
\centering
\caption{Stability of the glenohumeral joint, indicated as the angle [\si{\degree}], mean$\pm$SD~(max), between the JRF and the pure compression direction.}
\label{tab:stability}
\begin{tabular}{lccc}
\toprule
\emph{\small [\si{\degree}]}  & EAT     & WASH    & COMB    \\
\midrule
intact & 13$\pm4$(19) & 17$\pm3$(24) & 13$\pm$\phantom{0}9(32) \\
$-$SSC   & 30$\pm5$(39) & 23$\pm5$(37) & 14$\pm12$(36) \\
+PMAC  & 28$\pm6$(39) & 22$\pm6$(36) & 13$\pm12$(35) \\
+PMAS  & 24$\pm7$(39) & 22$\pm6$(36) & 14$\pm10$(36) \\
+PMASC & 25$\pm7$(39) & 22$\pm7$(37) & 11$\pm11$(35) \\
\bottomrule
\end{tabular}
\end{table}

\section{Discussion}
In this study, we report validations with three essential model outputs.
First, our model was capable of replicating the arm motion for the considered ADL with very limited positional error.
Second, the computed muscle activation generally matched positively with the EMG of the reference subject for a large set of muscles, showing that our model can predict realistic activation pattern.
Finally, the simulated joint reaction force conformed to in-vivo measurements. 
Further model validations can be made by comparisons to clinical observations below: 

First, the transfer of PMAC leads to a substantial increase in DAN activation, indicating that some load is  transferred from PMAC to DAN.
This model observation is in agreement with the clinical long-term post-operative observation of DAN excessively bulging after PMA transfers~\cite{Resch2000}. In addition, current surgical indications include a functioning deltoid as a pre-requisite for the PMA transfer~\cite{Nelson2014}.

Second, stability of the joint is not remarkably improved after transfer of PMA. This could explain the negative clinical outcome observed in patients suffering from anterior subluxation of the shoulder prior to the PMA transfer~\cite{Elhassan2008}, and may confirm that this transfer is unsuitable in cases of anterior instability.
Co-contraction of the ISP, an antagonist of SSC, is reduced for all motions in the torn SSC case, when compared to the intact case.
Such a reduction in co-contraction affects joint stability~\cite{Gasbarro2017}, by reducing the compressive JRF component.
We observed the PMA transfer to partially restore co-contraction for eating and combing motions.
During the combing motion, the SSC is mostly inactive but the observed effects of SSC tear on ISP co-contraction may be the result of passive forces by SSC tissue.

Third, a torn SSC induces significant changes in activation and loading of the SSP.
The higher SSP load may explain why SSC tears, when not treated, often propagate in time superiorly leading to SSP tears~\cite{Lenart2017,Lee2018subscapularis}.
Our simulated transfers led to a load reduction of the SSP compared to the torn SSC case, explained by the transferred PMA taking over the SSC function.
Therefore, an early diagnosis of such pathology and its repair with a muscle transfer might limit SSP overuse and hence the risk of tear propagation from an isolated tear into a massive rotator cuff pathology. This is especially important with such transfers having been shown to be less successful in patients suffering from massive tears~\cite{Jost2003,Nelson2014,Shin2016}.

Furthermore, the transferred muscle is typically assumed to activate in-phase with the intact SSC~\cite{Shin2016}. This assumption often leads to omitting any biofeedback exercises during rehabilitation~\cite{Omid2013}. 
Our simulations showed that PMAC and SSC may fulfill similar functions, but they do not act in-phase during complex movements, but rather complementary to one another depending on the activity.
The activation pattern of the transferred muscle changes significantly compared to its initial state. 
Therefore, the neurological adaptation necessary to cope with the new situation after a transfer may be challenging for some patients, as they need to re-learn a completely different neuromuscular control pattern.

In this study, we considered PMA transfers on a single subject, for a limited number of ADL. Subject variability may be a potential contributor in the observations and further studies are required to generalize the results of this work. 
While the three considered activities are common ADL covering relatively different ranges, they still do not cover the full range of shoulder motions. Simulating supplementary movements might provide additional insights.

Due to the absence of ground truth for the optimal rest stretch ratio in our muscles models, we assumed that the rest position for all muscles was at the standard anatomical position, whereas in reality, many muscles are not at their intrinsic optimal rest length in the standard anatomical position~\cite{Mount2003}.
This might, for example, have restricted the range of motion, especially over head, due to some muscles not stretching as much as physiologically possible. Alternatively, some muscles may also have been unable to contract as much as physiologically possible.
We indeed observed this for latissimus dorsi and teres major for the combing motion, which required an increased value of optimal rest stretch ratio compared to default value.
Our model does not simulate inter-muscles collisions, which prevents the evaluation of PMA transfers underneath or over the conjoint tendon, which would allow a stability vs.\ risk tradeoff~\cite{Nelson2014,Thompson2020}.

\section{Conclusions}
Muscle transfers have a significant impact on the neuromuscular system during active motions, beyond mere kinematics.
While the subscapularis and pectoralis major fulfill similar functions, they do not necessarily activate at similar moments during a motion. 
Activation patterns differ greatly after transfers compared to intact state, which may explain the inability of some patients to adapt after a PMA transfer.
In particular, rehabilitation exercises after transfer of the clavicular part of the pectoralis major should focus on washing and combing motions, which exhibit clear antagonistic patterns.
A transfer of the clavicular part of the pectoralis major alone, or together with the sternal part, may not be successful when the patient suffers from a weak anterior deltoid.
In general, the muscle transfers considered in this study were shown to reduce the load on the supraspinatus, decreasing the risks of tear propagation. The muscle transfers also restored antagonistic co-contraction of the infraspinatus.
Simulation of active musculoskeletal dynamics is shown to be an asset for assessing the outcome of muscle transfers. Such models may help in understanding the effect of tendon transfers on active muscle function and hence help guide the choice of the surgical approach. 

\section{Acknowledgments}
This work was funded partially by a Highly Specialized Medicine Grant from Canton of Zurich as well as the Surgent project of University Medicine Zurich.

\bibliographystyle{abbrvnat}
\bibliography{arxiv}

\begin{thebibliography}{43}
\providecommand{\natexlab}[1]{#1}
\providecommand{\url}[1]{\texttt{#1}}
\expandafter\ifx\csname urlstyle\endcsname\relax
  \providecommand{\doi}[1]{doi: #1}\else
  \providecommand{\doi}{doi: \begingroup \urlstyle{rm}\Url}\fi

\bibitem[Ackland et~al.(2008)Ackland, Pak, Richardson, and Pandy]{Ackland2008}
D.~C. Ackland, P.~Pak, M.~Richardson, and M.~G. Pandy.
\newblock {Moment arms of the muscles crossing the anatomical shoulder}.
\newblock \emph{Journal of Anatomy}, 213\penalty0 (4):\penalty0 383--390, 2008.

\bibitem[Axe(2016)]{Axe2016}
J.~M. Axe.
\newblock {Tendon transfers for irreparable rotator cuff tears: An update}.
\newblock \emph{EFORT Open Reviews}, 1\penalty0 (1):\penalty0 18--24, 2016.

\bibitem[Bergmann et~al.(2007)Bergmann, Graichen, Bender, K{\"{a}}{\"{a}}b,
  Rohlmann, and Westerhoff]{Bergmann2007}
G.~Bergmann, F.~Graichen, A.~Bender, M.~K{\"{a}}{\"{a}}b, A.~Rohlmann, and
  P.~Westerhoff.
\newblock {In vivo glenohumeral contact forces—Measurements in the first
  patient 7 months postoperatively}.
\newblock \emph{Journal of Biomechanics}, 40\penalty0 (10):\penalty0
  2139--2149, 2007.

\bibitem[Blemker et~al.(2005)Blemker, Pinsky, and Delp]{Blemker2005a}
S.~S. Blemker, P.~M. Pinsky, and S.~L. Delp.
\newblock {A 3D model of muscle reveals the causes of nonuniform strains in the
  biceps brachii}.
\newblock \emph{Journal of Biomechanics}, 38\penalty0 (4):\penalty0 657--665,
  2005.

\bibitem[Bolsterlee et~al.(2014)Bolsterlee, Veeger, and van~der
  Helm]{Bolsterlee2014b}
B.~Bolsterlee, H.~E.~J. Veeger, and F.~C.~T. van~der Helm.
\newblock {Modelling clavicular and scapular kinematics: from measurement to
  simulation}.
\newblock \emph{Medical {\&} Biological Engineering {\&} Computing},
  52\penalty0 (3):\penalty0 283--291, 2014.

\bibitem[Clark and Elhassan(2018)]{Clark2018}
N.~J. Clark and B.~T. Elhassan.
\newblock {The Role of Tendon Transfers for Irreparable Rotator Cuff Tears}.
\newblock \emph{Current Reviews in Musculoskeletal Medicine}, 11\penalty0
  (1):\penalty0 141--149, 2018.

\bibitem[Elhassan et~al.(2008)Elhassan, Ozbaydar, Massimini, Diller, Higgins,
  and Warner]{Elhassan2008}
B.~Elhassan, M.~Ozbaydar, D.~Massimini, D.~Diller, L.~Higgins, and J.~J.~P.
  Warner.
\newblock {Transfer of pectoralis major for the treatment of irreparable tears
  of subscapularis: Does it work?}
\newblock \emph{Journal of Bone and Joint Surgery - Series B}, 90\penalty0
  (8):\penalty0 1059--1065, 2008.

\bibitem[Elhassan et~al.(2010)Elhassan, Bishop, Shin, and
  Spinner]{Elhassan2010}
B.~Elhassan, A.~Bishop, A.~Shin, and R.~Spinner.
\newblock {Shoulder Tendon Transfer Options for Adult Patients With Brachial
  Plexus Injury}.
\newblock \emph{Journal of Hand Surgery}, 35\penalty0 (7):\penalty0 1211--1219,
  2010.

\bibitem[Elhassan et~al.(2014)Elhassan, Christensen, and Wagner]{Elhassan2014}
B.~Elhassan, T.~J. Christensen, and E.~R. Wagner.
\newblock {Feasibility of latissimus and teres major transfer to reconstruct
  irreparable subscapularis tendon tear: An anatomic study}.
\newblock \emph{Journal of Shoulder and Elbow Surgery}, 23\penalty0
  (4):\penalty0 492--499, 2014.

\bibitem[Ernstbrunner et~al.(2019)Ernstbrunner, Wieser, Catanzaro, Agten,
  Fornaciari, Bauer, and Gerber]{Ernstbrunner2019}
L.~Ernstbrunner, K.~Wieser, S.~Catanzaro, C.~A. Agten, P.~Fornaciari, D.~E.
  Bauer, and C.~Gerber.
\newblock {Long-Term Outcomes of Pectoralis Major Transfer for the Treatment of
  Irreparable Subscapularis Tears}.
\newblock \emph{The Journal of Bone and Joint Surgery}, 101\penalty0
  (23):\penalty0 2091--2100, 2019.

\bibitem[Favre et~al.(2008)Favre, Loeb, Helmy, and Gerber]{Favre2008a}
P.~Favre, M.~D. Loeb, N.~Helmy, and C.~Gerber.
\newblock {Latissimus dorsi transfer to restore external rotation with reverse
  shoulder arthroplasty: A biomechanical study}.
\newblock \emph{Journal of Shoulder and Elbow Surgery}, 17\penalty0
  (4):\penalty0 650--658, 2008.

\bibitem[Gasbarro et~al.(2017)Gasbarro, Bondow, and Debski]{Gasbarro2017}
G.~Gasbarro, B.~Bondow, and R.~Debski.
\newblock {Clinical anatomy and stabilizers of the glenohumeral joint}.
\newblock \emph{Annals of Joint}, 2:\penalty0 58--58, 2017.

\bibitem[Gavriilidis et~al.(2010)Gavriilidis, Kircher, Magosch, Lichtenberg,
  and Habermeyer]{Gavriilidis2010}
I.~Gavriilidis, J.~Kircher, P.~Magosch, S.~Lichtenberg, and P.~Habermeyer.
\newblock {Pectoralis major transfer for the treatment of irreparable
  anterosuperior rotator cuff tears}.
\newblock \emph{International Orthopaedics}, 34\penalty0 (5):\penalty0
  689--694, 2010.

\bibitem[Gerber and Krushell(1991)]{Gerber1991}
C.~Gerber and R.~J. Krushell.
\newblock {Isolated rupture of the tendon of the subscapularis muscle. Clinical
  features in 16 cases.}
\newblock \emph{The Journal of bone and joint surgery. British volume},
  73\penalty0 (3):\penalty0 389--394, 1991.

\bibitem[Gerber et~al.(1996)Gerber, Hersche, and Farron]{Gerber1996}
C.~Gerber, O.~Hersche, and A.~Farron.
\newblock {Isolated rupture of the subscapularis tendon.}
\newblock \emph{The Journal of bone and joint surgery. American volume},
  78\penalty0 (7):\penalty0 1015--23, 1996.

\bibitem[Gumina and Candela(2017)]{Gumina2017clinical}
S.~Gumina and V.~Candela.
\newblock {Clinical Evaluation}.
\newblock In \emph{Rotator Cuff Tear}, pages 139--162. Springer International
  Publishing, Cham, 2017.

\bibitem[Hik and Ackland(2019)]{Hik2019}
F.~Hik and D.~C. Ackland.
\newblock {The moment arms of the muscles spanning the glenohumeral joint: a
  systematic review.}
\newblock \emph{Journal of anatomy}, 234\penalty0 (1):\penalty0 1--15, 2019.

\bibitem[Jastifer et~al.(2012)Jastifer, Gustafson, Patel, and
  Uggen]{Jastifer2012}
J.~Jastifer, P.~Gustafson, B.~Patel, and C.~Uggen.
\newblock {Pectoralis Major Transfer for Subscapularis Deficiency: A
  Computational Study}.
\newblock \emph{Shoulder {\&} Elbow}, 4\penalty0 (1):\penalty0 25--29, 2012.

\bibitem[Jost et~al.(2003)Jost, Puskas, Lustenberger, and Gerber]{Jost2003}
B.~Jost, G.~J. Puskas, A.~Lustenberger, and C.~Gerber.
\newblock {Outcome of pectoralis major transfer for the treatment of
  irreparable subscapularis tears.}
\newblock \emph{The Journal of bone and joint surgery. American volume},
  85\penalty0 (10):\penalty0 1944--51, 2003.

\bibitem[Keating et~al.(1993)Keating, Waterworth, Shaw-Dunn, and
  Crossan]{Keating1993}
J.~Keating, P.~Waterworth, J.~Shaw-Dunn, and J.~Crossan.
\newblock {The relative strengths of the rotator cuff muscles. A cadaver
  study}.
\newblock \emph{The Journal of Bone and Joint Surgery. British volume},
  75-B\penalty0 (1):\penalty0 137--140, 1993.

\bibitem[Konrad et~al.(2007)Konrad, Sudkamp, Kreuz, Jolly, McMahon, and
  Debski]{Konrad2007}
G.~G. Konrad, N.~P. Sudkamp, P.~C. Kreuz, J.~T. Jolly, P.~J. McMahon, and R.~E.
  Debski.
\newblock {Pectoralis major tendon transfers above or underneath the conjoint
  tendon in subscapularis-deficient shoulders: An in vitro biomechanical
  analysis}.
\newblock \emph{Journal of Bone and Joint Surgery - Series A}, 89\penalty0
  (11):\penalty0 2477--2484, 2007.

\bibitem[Lee et~al.(2018{\natexlab{a}})Lee, Shukla, and
  S{\'{a}}nchez-Sotelo]{Lee2018subscapularis}
J.~Lee, D.~R. Shukla, and J.~S{\'{a}}nchez-Sotelo.
\newblock {Subscapularis tears: hidden and forgotten no more}.
\newblock \emph{JSES Open Access}, 2\penalty0 (1):\penalty0 74--83,
  2018{\natexlab{a}}.

\bibitem[Lee et~al.(2018{\natexlab{b}})Lee, Yu, Park, Aanjaneya, Sifakis, and
  Lee]{Lee2018}
S.~Lee, R.~Yu, J.~Park, M.~Aanjaneya, E.~Sifakis, and J.~Lee.
\newblock {Dexterous manipulation and control with volumetric muscles}.
\newblock \emph{ACM Transactions on Graphics}, 37\penalty0 (4):\penalty0 1--13,
  2018{\natexlab{b}}.

\bibitem[Lenart and Ticker(2017)]{Lenart2017}
B.~A. Lenart and J.~B. Ticker.
\newblock {Subscapularis tendon tears: Management and arthroscopic repair}.
\newblock \emph{EFORT Open Reviews}, 2\penalty0 (12):\penalty0 484--495, 2017.

\bibitem[Magermans et~al.(2004{\natexlab{a}})Magermans, Chadwick, Veeger,
  van~der Helm, and Rozing]{Magermans2004biomechanical}
D.~Magermans, E.~Chadwick, H.~Veeger, F.~van~der Helm, and P.~Rozing.
\newblock {Biomechanical analysis of tendon transfers for massive rotator cuff
  tears}.
\newblock \emph{Clinical Biomechanics}, 19:\penalty0 350--357,
  2004{\natexlab{a}}.

\bibitem[Magermans et~al.(2004{\natexlab{b}})Magermans, Chadwick, Veeger,
  van~der Helm, and Rozing]{Magermans2004effectiveness}
D.~Magermans, E.~Chadwick, H.~Veeger, F.~van~der Helm, and P.~Rozing.
\newblock {Effectiveness of tendon transfers for massive rotator cuff tears : a
  simulation study}.
\newblock \emph{Clinical Biomechanics}, 19:\penalty0 116--122,
  2004{\natexlab{b}}.

\bibitem[Minagawa et~al.(2013)Minagawa, Yamamoto, Abe, Fukuda, Seki, Kikuchi,
  Kijima, and Itoi]{Minagawa2013}
H.~Minagawa, N.~Yamamoto, H.~Abe, M.~Fukuda, N.~Seki, K.~Kikuchi, H.~Kijima,
  and E.~Itoi.
\newblock {Prevalence of symptomatic and asymptomatic rotator cuff tears in the
  general population: From mass-screening in one village}.
\newblock \emph{Journal of Orthopaedics}, 10\penalty0 (1):\penalty0 8--12,
  2013.

\bibitem[Moroder et~al.(2017)Moroder, Schulz, Mitterer, Plachel, Resch, and
  Lederer]{Moroder2017}
P.~Moroder, E.~Schulz, M.~Mitterer, F.~Plachel, H.~Resch, and S.~Lederer.
\newblock {Long-term outcome after pectoralis major transfer for irreparable
  anterosuperior rotator cuff tears}.
\newblock \emph{Journal of Bone and Joint Surgery - American Volume},
  99\penalty0 (3):\penalty0 239--245, 2017.

\bibitem[Mount et~al.(2003)Mount, Whitmore, and Stealey]{Mount2003}
F.~Mount, M.~Whitmore, and S.~Stealey.
\newblock {Evaluation of neutral body posture on shuttle mission sts-57
  (spacehab-1)}.
\newblock \emph{National Aeronautics and Space Administration (NASA),
  Washington DC}, 57\penalty0 (February), 2003.

\bibitem[Nelson et~al.(2014)Nelson, Namdari, Galatz, and Keener]{Nelson2014}
G.~N. Nelson, S.~Namdari, L.~Galatz, and J.~D. Keener.
\newblock {Pectoralis major tendon transfer for irreparable subscapularis
  tears}.
\newblock \emph{Journal of Shoulder and Elbow Surgery}, 23\penalty0
  (6):\penalty0 909--918, 2014.

\bibitem[Omid and Lee(2013)]{Omid2013}
R.~Omid and B.~Lee.
\newblock {Tendon transfers for irreparable rotator cuff tears.}
\newblock \emph{The Journal of the American Academy of Orthopaedic Surgeons},
  21\penalty0 (8):\penalty0 492--501, 2013.

\bibitem[Paladini et~al.(2013)Paladini, Campi, Merolla, Pellegrini, and
  Porcellini]{Paladini2013}
P.~Paladini, F.~Campi, G.~Merolla, A.~Pellegrini, and G.~Porcellini.
\newblock {Pectoralis minor tendon transfer for irreparable anterosuperior cuff
  tears}.
\newblock \emph{Journal of Shoulder and Elbow Surgery}, 22\penalty0
  (6):\penalty0 e1, 2013.

\bibitem[{P{\'e}an} et~al.(2020){P{\'e}an}, {Favre}, and {Goksel}]{Pean2020rsa}
F.~{P{\'e}an}, P.~{Favre}, and O.~{Goksel}.
\newblock {Influence of Rotator Cuff Integrity on Loading and Kinematics Before
  and After Reverse Shoulder Arthroplasty}.
\newblock \emph{arXiv e-prints}, art. arXiv:2012.09763, 2020.

\bibitem[Péan and Goksel(2020)]{Pean2020surface}
F.~Péan and O.~Goksel.
\newblock Surface-based modeling of muscles: Functional simulation of the
  shoulder.
\newblock \emph{Medical Engineering \& Physics}, 2020.

\bibitem[Resch et~al.(2000)Resch, Povacz, Ritter, and Matschi]{Resch2000}
H.~Resch, P.~Povacz, E.~Ritter, and W.~Matschi.
\newblock {Transfer of the pectoralis major muscle for the treatment of
  irreparable rupture of the subscapularis tendon}.
\newblock \emph{Journal of Bone and Joint Surgery - Series A}, 82\penalty0
  (3):\penalty0 372--382, 2000.

\bibitem[Shin et~al.(2016)Shin, Saccomanno, Cole, Romeo, Nicholson, and
  Verma]{Shin2016}
J.~J. Shin, M.~F. Saccomanno, B.~J. Cole, A.~A. Romeo, G.~P. Nicholson, and
  N.~N. Verma.
\newblock {Pectoralis major transfer for treatment of irreparable subscapularis
  tear: a systematic review}.
\newblock \emph{Knee Surgery, Sports Traumatology, Arthroscopy}, 24\penalty0
  (6):\penalty0 1951--1960, 2016.

\bibitem[Staudenmann et~al.(2007)Staudenmann, Potvin, Kingma, Stegeman, and van
  Die{\"{e}}n]{Staudenmann2007}
D.~Staudenmann, J.~R. Potvin, I.~Kingma, D.~F. Stegeman, and J.~H. van
  Die{\"{e}}n.
\newblock {Effects of EMG processing on biomechanical models of muscle joint
  systems: Sensitivity of trunk muscle moments, spinal forces, and stability}.
\newblock \emph{Journal of Biomechanics}, 40\penalty0 (4):\penalty0 900--909,
  2007.

\bibitem[Thompson et~al.(2020)Thompson, Kwon, Flatow, Jazrawi, Strauss, and
  Alaia]{Thompson2020}
K.~Thompson, Y.~Kwon, E.~Flatow, L.~Jazrawi, E.~Strauss, and M.~Alaia.
\newblock {Everything pectoralis major: from repair to transfer}.
\newblock \emph{Physician and Sportsmedicine}, 48\penalty0 (1):\penalty0
  33--45, 2020.

\bibitem[Valenti et~al.(2011)Valenti, Kany, Ferriere, Kilinc, and
  Thomsen]{Valenti2011}
P.~Valenti, J.~Kany, S.~Ferriere, A.~Kilinc, and L.~Thomsen.
\newblock {Transfer of the pectoralis major for the treatment of irreparable
  subscapularis tear: review of 15 cases}.
\newblock In \emph{Tendon transfer for irreparable cuff tear}, number~14, pages
  95--110. Springer Paris, Paris, 2011.

\bibitem[Valenti et~al.(2015)Valenti, Boughebri, Moraiti, Dib, Maqdes, Amouyel,
  Ciais, and Kany]{Valenti2015}
P.~Valenti, O.~Boughebri, C.~Moraiti, C.~Dib, A.~Maqdes, T.~Amouyel, G.~Ciais,
  and J.~Kany.
\newblock {Transfer of the clavicular or sternocostal portion of the pectoralis
  major muscle for irreparable tears of the subscapularis. Technique and
  clinical results}.
\newblock \emph{International Orthopaedics}, 39\penalty0 (3):\penalty0
  477--483, 2015.

\bibitem[Westerhoff et~al.(2009)Westerhoff, Graichen, Bender, Halder, Beier,
  Rohlmann, and Bergmann]{Westerhoff2009a}
P.~Westerhoff, F.~Graichen, A.~Bender, A.~Halder, A.~Beier, A.~Rohlmann, and
  G.~Bergmann.
\newblock {In vivo measurement of shoulder joint loads during activities of
  daily living}.
\newblock \emph{Journal of Biomechanics}, 42\penalty0 (12):\penalty0
  1840--1849, 2009.

\bibitem[Wirth and Rockwood(1997)]{Wirth1997}
M.~A. Wirth and C.~A. Rockwood.
\newblock {Operative treatment of irreparable rupture of the subscapularis.}
\newblock \emph{The Journal of bone and joint surgery. American volume},
  79\penalty0 (5):\penalty0 722--31, 1997.

\bibitem[Wu et~al.(2005)Wu, {Van Der Helm}, Veeger, Makhsous, {Van Roy},
  Anglin, Nagels, Karduna, McQuade, Wang, Werner, and Buchholz]{Wu2005}
G.~Wu, F.~C. {Van Der Helm}, H.~Veeger, M.~Makhsous, P.~{Van Roy}, C.~Anglin,
  J.~Nagels, A.~R. Karduna, K.~McQuade, X.~Wang, F.~W. Werner, and B.~Buchholz.
\newblock {ISB recommendation on definitions of joint coordinate systems of
  various joints for the reporting of human joint motion - Part II: Shoulder,
  elbow, wrist and hand}.
\newblock \emph{Journal of Biomechanics}, 38\penalty0 (5):\penalty0 981--992,
  2005.

\end{thebibliography}
\end{document}